\documentclass[]{aastex631}

\begin{document}

\title{Using the COSMIC Population Synthesis Code to Investigate How Metallicity Affects the Rates of Interacting Binaries}

\author[0009-0004-6379-2460]{Ayanah L Cason}
\affiliation{Computational Physics and Methods Group (CCS-2), Los Alamos National Laboratory; Los Alamos, NM 87544}

\author[0000-0003-1707-7998]{Nicole Lloyd-Ronning }
\affiliation{Computational Physics and Methods Group (CCS-2), Los Alamos National Laboratory; Los Alamos, NM 87544}

\author[0000-0002-4854-8636]{Roseanne Cheng}
\affiliation{Computational Physics and Methods Group (CCS-2), Los Alamos National Laboratory; Los Alamos, NM 87544}
\affiliation{Fluid Dynamics and Solid Mechanics Group (T-3), Los Alamos National Laboratory; Los Alamos, NM 87544}



\begin{abstract}

 We use COSMIC, a galaxy population synthesis code, to investigate how metallicity affects the rate of formation of massive stars with a closely orbiting compact object companion, the suggested progenitors of radio loud long gamma-ray bursts. We present the evolution time of these systems at different metallicities, and how the formation rates of these systems are anti-correlated with metallicity.  In particular, these systems occur about 10 times more frequently in at metallicities between $Z = 2\times 10^{-4}$ and $2 \times 10^{-3}$, compared to those between $Z = 2\times 10^{-3}$ and $2 \times 10^{-2}$.  This work serves as a prerequisite to predicting the global rates of these systems as a function of redshift, ultimately giving crucial insight into our understanding of the progenitors of long gamma-ray bursts and their evolution over cosmic time.

\end{abstract}

\keywords{Gamma-ray Bursts (629)--- Binary Stars (154) --- Astronomical Simulations (1857) --- Metallicity(1031)}


\section{Introduction} \label{sec:intro}
The progenitors of gamma-ray bursts - the brightest explosions in our universe - are still poorly understood. It has been suggested that massive stars collapsing in interacting
binary systems (i.e. with a closely orbiting compact object companion) can provide the necessary angular momentum and dense surrounding medium to explain observations of radio-loud gamma-ray bursts \citep{LR22}. We explore the dependence of the formation rates of such interacting binary systems on metallicity. 
Metallicity - defined as the proportion of elements heavier than helium present in a system - is a key physical quantity that evolves as the universe evolves, increasing as redshift decreases (as more stars are formed throughout the history of the universe, they synthesize more metals which then become the ingredients for subsequent generations of stars). This quantity influences star formation (including binary and multiple formation) and stellar evolution, and it is important to understand how metallicity affects the formation of stellar systems throughout cosmic time.\\

Exactly how metallicity affects binary evolution is still very much an open question.  \cite{Machida2009} conclude that there is an anticorrelation between the rates of binaries star systems and metallicity.  
\cite{Klencki2020} show that lower metallicities in binary systems cause the massive star to become stripped of its envelope later than those of higher metallicity, which allows the star to have a more massive core. These two components allow more mass transfer before its collapse, ultimately affecting the final state of the system. \cite{Ponnada2020} also conclude that the rate of high mass x-ray binaries (HMXB) increases as the metallicity decreases. \cite{Boesky24} show that binary black hole (BBH) rates increase with decreasing metallicity. Finally, \cite{Boesky2024} show that the deviation of the BBH rate from the star formation rate could be due to BBH rate's dependence on metallicity. The objective of this work is to contribute to the understanding of metallicity dependence specifically in systems that consist of a massive star and a closely orbiting compact companion. \\

This work uses population synthesis to investigate how the rates of these systems (a massive star with a closely orbiting compact object companion) change with metallicity. This paper is organized as follows: In \S \ref{sec:Methods}, we describe the code we use and the post-processing procedures we apply to obtain our results. In \S \ref{sec:Results} we present our results. In \S \ref{sec:Conclusions} we present our conclusions and discuss the implications of this work.\\\\

\section{Methods} \label{sec:Methods}

We use the open source population synthesis code COSMIC \citep{Breivik2020} to investigate the rates of massive stars that have a closely orbiting compact object companion.  COSMIC is based on equations from \cite{Hurley2000} and \cite{Hurley2002} to model the evolutionary track of single-star formation and binary-star formation, respectively. COSMIC samples user-specified convergence criteria. Our convergence criteria are described below. For all other parameters, we use the default values in COSMIC, as described in \cite{Breivik2020}.\\

The binary type in which we are interested is a massive star with a closely orbiting black hole companion, where the tidal interaction between the two bodies leads to spin up of the massive star. We select massive stars in a mass range of 15 to 50 $M_{\odot}$, with a black hole companion (whose mass ranges from about 10 to 15 $M_{\odot}$, although we do not constrain this mass), and an orbital period ranging from 1.5 to 6 days.  It has been shown \cite{Hern24} that these systems are excellent candidates for GRB progenitors (where the GRB is created when the massive star collapses), given the angular momentum of the star at the end of its life. 
We explore the formation rate of these systems over a metallicity range of $10^{-4}<Z<3 \times 10^{-2}$.\\


To process and normalize the converged binaries, we divide the total number of converged systems by the total number of stars in the galaxy COSMIC simulates. 
\begin{equation}
    N_{IB}= N_{conv}/N_{stars}
\end{equation}
Where $N_{stars}$ is total number of stars in the simulation system (including single stars).


\begin{figure}
    \centering
    \includegraphics[width=18cm,height=7.5cm]{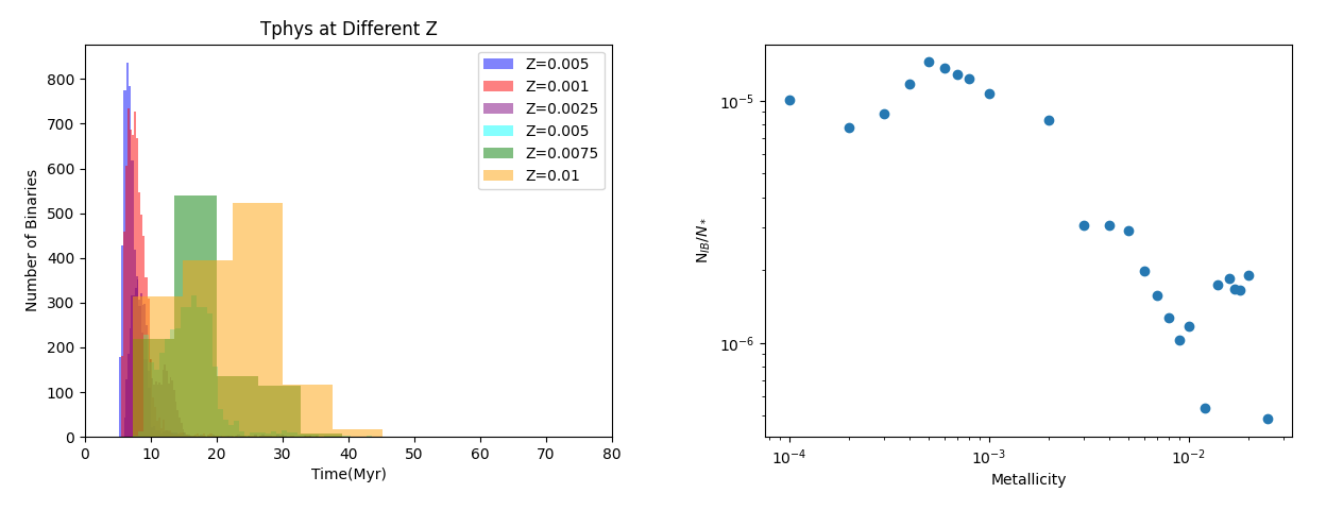}
    \caption{{\bf Left Panel:} Evolution time of our selected binary systems (see section \ref{sec:Methods} for our selection criteria) for six different metallicities. {\bf Right Panel:} Number of interacting binary systems composed of a massive star with a black hole companion divided by the total number of stars in the simulation vs. metallicity. }
    \label{fig:combo}
\end{figure}



\section{Results}\label{sec:Results}

The left panel of Figure \ref{fig:combo} shows the evolution time of the binary systems for six different metallicities. The evolution time distribution is narrower at lower metallicities, with a peak at $\sim 7$ Myr. At higher metallicities (as expected due to the more complex chemical make-up of the environment and the stars themselves) the evolution times range from $\sim 10$ to $40$ Myr.  We use these results to help inform our convergence tests and the initial conditions of our subsequent simulations - in particular, where to set the ``start'' time of the simulation such that we allow sufficient time for the binary to form and evolve.  \\ 

The right panel of Figure \ref{fig:combo} shows the number of interacting binaries as a function of metallicity. We see these systems form more readily and abundantly at a lower metallicity.  In \cite{Breivik2020}, it is emphasized that different parts of a galaxy (bulge, thin disk, thick disk) have in general different values for their metallicities. For instance, \cite{Breivik2020} show that the thin and thick disks for a galaxy have more metal rich stars than the bulge. Given a distribution of metallicities throughout a given galaxy, the right panel of Figure \ref{fig:combo} can be used as a tool to constrain the  parts of the galaxy in which these systems may have a higher rate of occurrence. \\



There are two peaks in the right panel of Figure \ref{fig:combo}. The first, most prominent peak is at lower metallicity around $Z \sim 5 \times 10^{-4}$. There is another small peak around solar metallicity ($Z \sim 1.7 \times 10^{-2}$). Because most of the COSMIC code parameters were calibrated to near solar metallicity, we suggest the small, secondary peak in the right panel of Figure \ref{fig:combo} may be a numerical feature outside the range of interpretation. While it is beyond the scope of this paper, comparisons with other populuations synthesis codes can be used to check this statement.

\section{Conclusion}\label{sec:Conclusions}
We have shown the evolution times and relative fraction of massive stars with a closely orbiting compact object companion over a range of metallicities, using a suite of simulations from the population synthesis code COSMIC.  We find that the average evolution time of these systems, regardless of metallicity, is under 50 Myr, but that lower metallicity systems have a narrower distribution in their formation time and on average shorter evolution times. We show that a higher fraction of these systems form at low metallicity compared to higher  metallicity by about an order of magnitude.  This work helps set the groundwork for predicting the rates of these systems over cosmic time, with implications for understanding the progenitors of GRBs (particularly radio loud GRBs) and interpreting the observed evolution of GRB properties with redshift.  Because some of these systems will also eventually evolve to black hole binaries (if they are not disrupted when the massive star dies), we can also use this work to help us understand the contribution of these systems to the global stellar black hole binary merger rate, detectable by LIGO/Virgo/Kagra \citep{Wempe2024, Boesky2024} and future gravitational wave facilities.

\section{Acknowledgements}
\begin{acknowledgments}
We thank Katie Breivik, Mike Zevin, Shane Larson, Gabriel Casabona, and the COSMIC community for helpful conversations and the many resources provided to help run COSMIC.   
This work was supported by the U.~S. Department of Energy through Los Alamos National Laboratory (LANL).  LANL is operated by Triad National Security, LLC, for the National Nuclear Security Administration of U.S. Department of Energy (Contract No. 89233218CNA000001).   Research presented was supported by the Laboratory Directed Research and Development program of LANL project number 20230115ER. We acknowledge LANL Institutional Computing HPC Resources under project w23extremex. Additional research presented in this article was supported by the 
Laboratory Directed Research and Development program of Los Alamos National Laboratory under 
project number 20210808PRD1. LA-UR
\end{acknowledgments}

%

\vspace{5mm}





\bibliography{refs}{}
\bibliographystyle{aasjournal}



\end{document}